\documentclass[usenatbib]{mn2e}
\newcommand{\subL}{_{_{\!\ell\!o\!g\!M}}}
\newcommand{\subO}{_{_{\rm O}}}
\newcommand{\subS}{_{_{\rm S}}}
\usepackage{graphicx,tabularx,amsmath,amssymb,upgreek,wasysym,mathtools,multirow}
\numberwithin{equation}{section}
\title[Star formation triggered by cloud-cloud collisions]{Star Formation triggered by cloud-cloud collisions}
\author[S. K. Balfour, A. P. Whitworth, D. A. Hubber, S. E. Jaffa]
{S. K. Balfour$^{1}$\thanks{E-mail: Scott.Balfour@astro.cf.ac.uk}, 
A. P. Whitworth$^{1}$, 
D. A. Hubber$^{2,3}$, 
S. E. Jaffa$^{1}$\\
$^{1}$School of Physics and Astronomy, Cardiff University, Cardiff CF24 3AA, Wales, UK\\
$^{2}$University Observatory Munich, Ludwig-Maximilians-University Munich, Scheinerstr.1, 81679 Munich, Germany \\
$^{3}$Excellence Cluster Universe, Boltzmannstr. 2, 85748 Garching, Germany}

\begin{document}
\pagerange{\pageref{firstpage}--\pageref{lastpage}} \pubyear{2013}
\maketitle
\label{firstpage}

%%%%%%%
\begin{abstract} 
We present the results of SPH simulations in which two clouds, each having mass $M\subO\!=\!500\,{\rm M}_{_\odot}$ and radius $R\subO\!=\!2\,{\rm pc}$, collide head-on at relative velocities of $\Delta v\subO =2.4,\;2.8,\;3.2,\;3.6\;{\rm and}\;4.0\,{\rm km}\,{\rm s}^{-1}$. There is a clear trend with increasing $\Delta v\subO$. At low $\Delta v\subO$, star formation starts later, and the shock-compressed layer breaks up into an array of predominantly radial filaments; stars condense out of these filaments and fall, together with residual gas, towards the centre of the layer, to form a single large-$N$ cluster, which then evolves by competitive accretion, producing one or two very massive protostars and a diaspora of ejected (mainly low-mass) protostars; the pattern of filaments is reminiscent of the hub and spokes systems identified recently by observers. At high $\Delta v\subO$, star formation occurs sooner and the shock-compressed layer breaks up into a network of filaments; the pattern of filaments here is more like a spider's web, with several small-$N$ clusters forming independently of one another, in cores at the intersections of filaments, and since each core only spawns a small number of protostars, there are fewer ejections of protostars. As the relative velocity is increased, the {\it mean} protostellar mass increases, but the {\it maximum} protostellar mass and the width of the mass function both decrease. We use a Minimal Spanning Tree to analyse the spatial distributions of protostars formed at different relative velocities.
\end{abstract}
%%%%%%%


%%%%%%%%%
\section{Introduction}
%%%%%%%%%

It is commonly assumed that collisions between interstellar clouds trigger star formation, by producing dense, gravitationally unstable layers of gas. However, it is difficult to evaluate the contribution that collisions make to the overall galactic star formation rate, as we do not know how frequently such collisions occur, nor how star formation proceeds when they do \citep[e.g][]{1992MNRAS.256..291P,Chapman1992r,Whitworth1994d,Heitsch2006a,Inoue2013ag,Dobbs2015c}. Moreover, instances where on-going star formation may have been triggered by cloud-cloud collisions are hard to identify, and the observations are difficult to interpret unambiguously \citep{Haworth2015a}. Possible instances include Westerlund 2 \citep{Furukawa2009,Ohama2010}, the Trifid Nebula \citep{Torii2011}, and the Serpens Main Cluster \citep{Duarte-Cabral2011}.

In this paper we present simulations of cloud-cloud collisions, and discuss the statistical properties and spatial distributions of the protostars that form. We limit consideration to head-on collisions involving a single cloud mass and a single cloud radius, and focus our attention on the consequences of varying the velocity of the collision, $v\subO$. The consequences of varying the cloud mass, the cloud radius, and the impact parameter of the collision, will be explored in subsequent papers.

In Section \ref{SEC:SSU} we describe the simulation set-up. In Section \ref{SEC:RES} we present and discuss the results. The main conclusions are summarised in Section \ref{SEC:CONC}.

%%%%%%%%%
\section{Simulation set-up}\label{SEC:SSU}
%%%%%%%%%

%%%%%%%%%%%%
\subsection{Initial conditions}
%%%%%%%%%%%%

Initially, the two colliding clouds both have mass $M\subO =500\,{\rm M}_{_\odot}$, radius $R\subO =2\,{\rm pc}$, uniform density $\rho\subO =1.01\times 10^{-21}\,{\rm g}\,{\rm cm}^{-3}$, and hence freefall time $t_{_{\rm FF}}=2.1\,{\rm Myr}$. The fraction of hydrogen, by mass, is $X = 0.70$, and the hydrogen is assumed to be molecular, so the initial molecular hydrogen number-density is $n_{_{{\rm H}_2}}= 2.1\times 10^2\,{\rm H}_2\,{\rm cm}^{-3}$. The gas is initially isothermal at $T\subO = 10\,{\rm K}$, corresponding to an isothermal sound speed $a\subO = 0.19\,{\rm km}\,{\rm s}^{-1}$. The cloud centres are initially at
\begin{eqnarray}
(x,y,z)&=&(\pm 2.02,0,0)\,{\rm pc}\,,
\end{eqnarray}
i.e. the clouds are almost touching, and the initial bulk-velocities of the clouds are
\begin{eqnarray}
(v_x,v_y,v_z)=(\mp v\subO ,0,0)\,.
\end{eqnarray}
{ We presume that the two colliding clouds are part of a larger, approximately virialised cloud complex, with mass between $4M\subO\!=\!2,000\,{\rm M}_{_\odot}$ and $40M\subO\!=\!20,000\,{\rm M}_{_\odot}$ -- and hence, using Larson's scaling relation \citep{Larson1981n}, velocity dispersion between $\sim 1.2\,{\rm km}\,{\rm s}^{-1}$ and $\sim 2.0\,{\rm km}\,{\rm s}^{-1}$. Consequently the clouds are given velocities $v\subO=1.2,\,1.4,\,1.6,\,1.8\;{\rm and}\;2.0\,{\rm km}\,{\rm s}^{-1}$, and they collide with relative velocities in the range $\Delta v\subO\!\sim\!2.4\,{\rm km}\,{\rm s}^{-1}$ to $\Delta v\subO\!\sim\!4.0\,{\rm km}\,{\rm s}^{-1}$. We discuss at the end of Section \ref{SEC:OVE} what happens if $\Delta v\subO$ falls outside this range. The Mach Numbers of the accretion shocks bounding the shock-compressed layer are of order 6 to 10, giving compression factors in the range 36 to 100.}

%%%%%%%%%%%%
\subsection{Constitutive physics}
%%%%%%%%%%%%

The gas is evolved with a barotropic equation of state,
\begin{eqnarray}
T(\rho)&=&T\subO\,\left\{1\,+\,\left(\!\frac{\rho}{\rho_{_{\rm CRIT}}}\!\right)^{\!2/3}\right\},
\end{eqnarray}
with $T\subO =10\,{\rm K}$ and $\rho_{_{\rm CRIT}} =10^{-14}\,{\rm g}\,{\rm cm}^{-3}$. At low-densities, $\rho\ll\rho_{_{\rm CRIT}}$, the gas is approximately isothermal at $\,\sim\! 10\,{\rm K}$. At high densities, $\rho\gg\rho_{_{\rm CRIT}}$, the gas is approximately adiabatic with adiabatic index $\gamma\simeq 5/3$ (corresponding to cool molecular hydrogen in which the rotational degrees of freedom are not significantly excited, plus some monatomic helium). This prescription mimics approximately the variation of temperature with density at the centre of a $1\,{\rm M}_{_\odot}$ protostar \citep{Masunaga1999}.

%%%%%%%%%%%%
\subsection{Numerical method}
%%%%%%%%%%%%

The simulations are performed with the Smoothed Particle Hydrodynamics code {\sc gandalf} (Hubber et al., in preparation). A binary tree is used to compute gravitational forces and collate particle neighbour lists. The grad-$h$ formulation of  the evolution equations is invoked \citet{Price2004}, and they are solved with a second-order leapfrog integrator. Artificial viscosity is treated using the method of \citet{Morris1997}, with parameters $\alpha_{_{min}} =0.1$, $\alpha =1$ and $\beta =2$. Sinks are introduced and evolved using the standard procedure described in \citet{Hubber2013}, with $\rho_{_{\rm SINK}}=10^{-12}\,{\rm g}\,{\rm cm}^{-3}$; hence condensations which are converted into sinks are already well into their Kelvin-Helmholtz contraction phase, and their properties are essentially independent of the choice of $\rho_{_{\rm SINK}}$. 

All simulations are performed with ${\cal N}_{_{\rm TOT}}\simeq10^6$ SPH particles, so each SPH particle has mass $m_{_{\rm SPH}}\simeq 0.001\,{\rm M}_{_\odot}$. The resolution-parameter is set to $\eta =1.2$, so that an SPH particle typically has $\,\sim\!57$ neighbours. Hence the mass resolution is $\sim 0.1\,{\rm M}_{_\odot}$ (i.e. just above the hydrogen-burning limit). The state of the simulation is output every $10,000\,{\rm yr}$.

Feedback from protostars is not included, and therefore the simulations are terminated once $10\%$ of the total mass has been converted into protostars, on the assumption that feedback would be terminating star formation around this stage. { To include feedback would require us (a) to reduce the resolution (increased $m_{_{\rm SPH}}$), with the consequence that low-mass H-burning stars could not form, and (b) to introduce sinks at a lower density threshold (lower $\rho_{_{\rm SINK}}$), so that the simulations could be followed further and massive stars ($>8\,{\rm M}_{_\odot}$) could form \citep[cf.][]{Dale2015}. We are presently repeating the simulations presented here with larger $m_{_{\rm SPH}}$ and lower $\rho_{_{\rm SINK}}$, with a view to understanding the formation and evolution of bipolar H{\sc ii} regions \citep[e.g.][]{Deharveng2015}, and these simulations will be presented in a future paper.}

%%%%
\begin{figure*}
\includegraphics[scale=0.26]{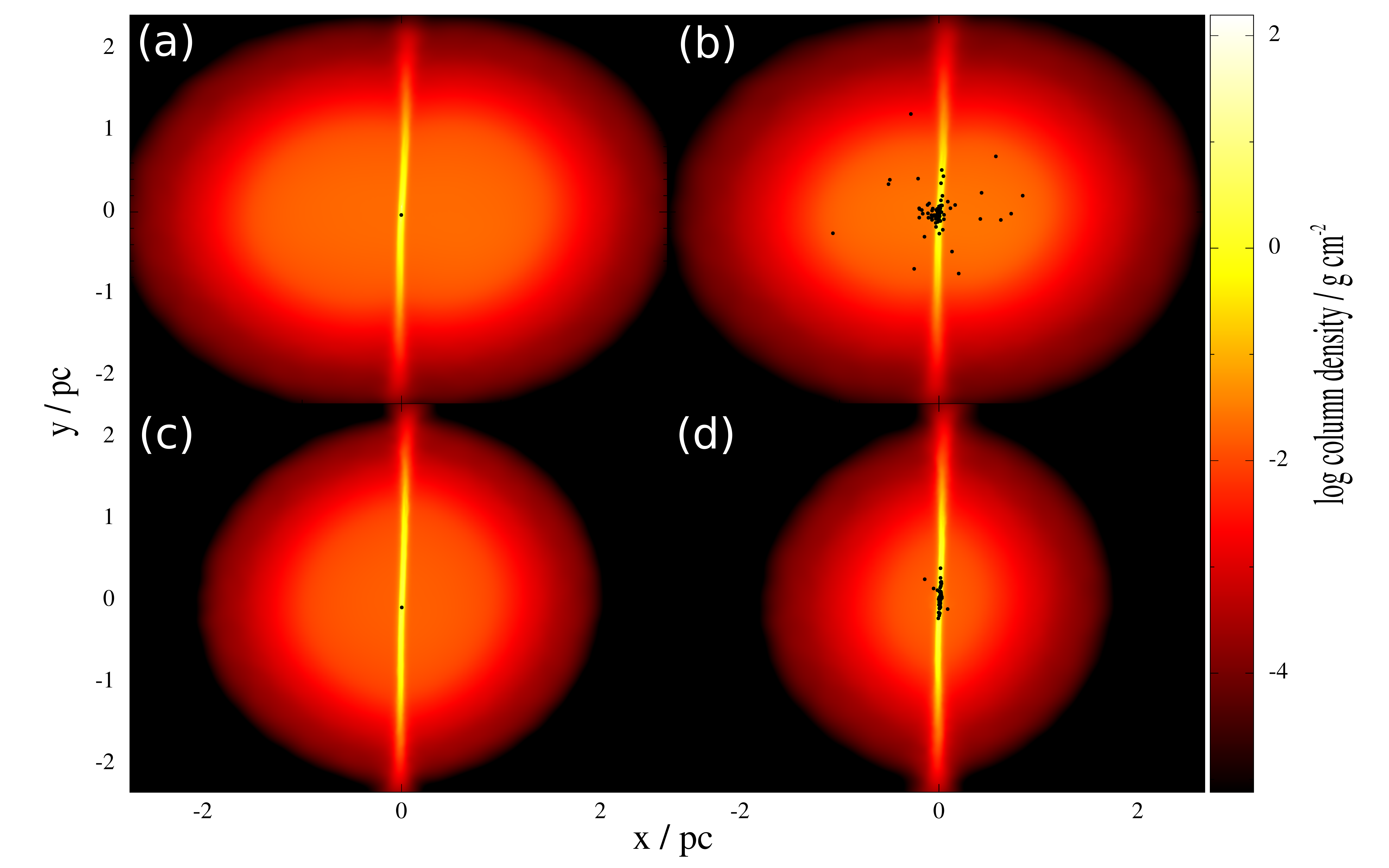}
\caption{{ False-colour images of colliding clouds, as seen looking across the collision axis, at $t_{_{\star 1}}$ (i.e. when the first protostar forms) and $t_{_{10\%}}$ (i.e. when $10\%$ of the mass has been converted into protostars). (a) $v\subO = 1.2\,{\rm km}\,{\rm s}^{-1}$, at $t_{_{\star 1}}=1.33\,{\rm Myr}$; (b) $v\subO = 1.2\,{\rm km}\,{\rm s}^{-1}$, at $t_{_{10\%}}=1.47\,{\rm Myr}$; (c) $v\subO = 2.0\,{\rm km}\,{\rm s}^{-1}$, at $t_{_{\star 1}}=1.14\,{\rm Myr}$; (d) $v\subO = 2.0\,{\rm km}\,{\rm s}^{-1}$, at $t_{_{10\%}}=1.28\,{\rm Myr}$.} Colour represents $\log_{_{10}}\!\left\{\Sigma/{\rm g}\,{\rm cm}^{-2}\right\}$, where $\Sigma$ is column-density projected on the $(x,y)$-plane. The black dots are protostars.}
\label{FIG:SideView}
\end{figure*}
%%%%

%%%%%%%%%%%%
\subsection{Perturbations}\label{SEC:PERTS}
%%%%%%%%%%%%

For each velocity, $v\subO$, we perform ten simulations. Of these ten simulations, five are performed with clouds that initially have zero internal velocity dispersion (Set 1), and five are performed with clouds that initially have a very low-amplitude internal turbulent velocity field (Set 2). 

For the Set 1 simulations, the SPH particles are given random positions in a cube with side $L_{_{\rm CUBE}}\!=\!4\,{\rm pc}$. The cube is then settled, using periodic boundary conditions, until the fractional density fluctuations have standard deviation less than $0.4\%$. Finally, particles are culled if they fall outside a sphere of radius $2\,{\rm pc}$. This sphere, and a clone of it rotated through random Euler angles, then represent the two clouds. The particles in a cloud have zero velocity dispersion. Initial density perturbations derive solely from the very small residual fluctuations in the particle density. The ratio of thermal to gravitational energy in an individual cloud is $\alpha_{_{\rm THERMAL}}\!=\!0.083\,<\,0.5$, so the pre-collision clouds are not virialised.

For the Set 2 simulations, the particles are distributed in the same way as Set 1. Next a random Gaussian turbulent velocity field, with power-spectrum $P_k\propto k^{-4}$, is generated on a $128^3$ grid, and the velocities of the SPH particles are computed by interpolation on this grid. Finally, the particle velocities are shifted to the centre of mass frame, and re-scaled so that they have a one-dimensional velocity dispersion $\sigma_{_{\rm TURB}}=0.066\,{\rm km}\,{\rm s}^{-1}$, corresponding to Mach Number ${\cal M}_{_{\rm TURB}}=0.35$. The ratio of turbulent to gravitational energy in an individual cloud is then $\alpha_{_{\rm TURB}}\!=\!0.010$, so the pre-collision clouds are not virialised, $\,\alpha_{_{\rm THERM}}+\alpha_{_{\rm TURB}}\,=\,0.093\,<\,0.5$.

The initial conditions are not intended to be realistic. Realistic clouds would not have uniform density, and they would have a much higher level of turbulence. The initial conditions are chosen because we want to isolate the effects of the collision from the effects of pre-existing density or velocity sub-structures. Also, we want to keep the parameter space as small as possible, in order to be able to quantify cause and effect. { We will explore in a future publication the extent to which the coherent large-scale structures generated here survive when the individual clouds have pre-existing density and velocity sub-structures of different amplitudes.

If evolved in isolation, the individual clouds collapse and fragment to form protostars, but they do so on a somewhat longer timescale than we consider here ($\ga 2\,{\rm Myr}$).}

%%%%%%%
\begin{figure*}
\includegraphics[scale=0.28]{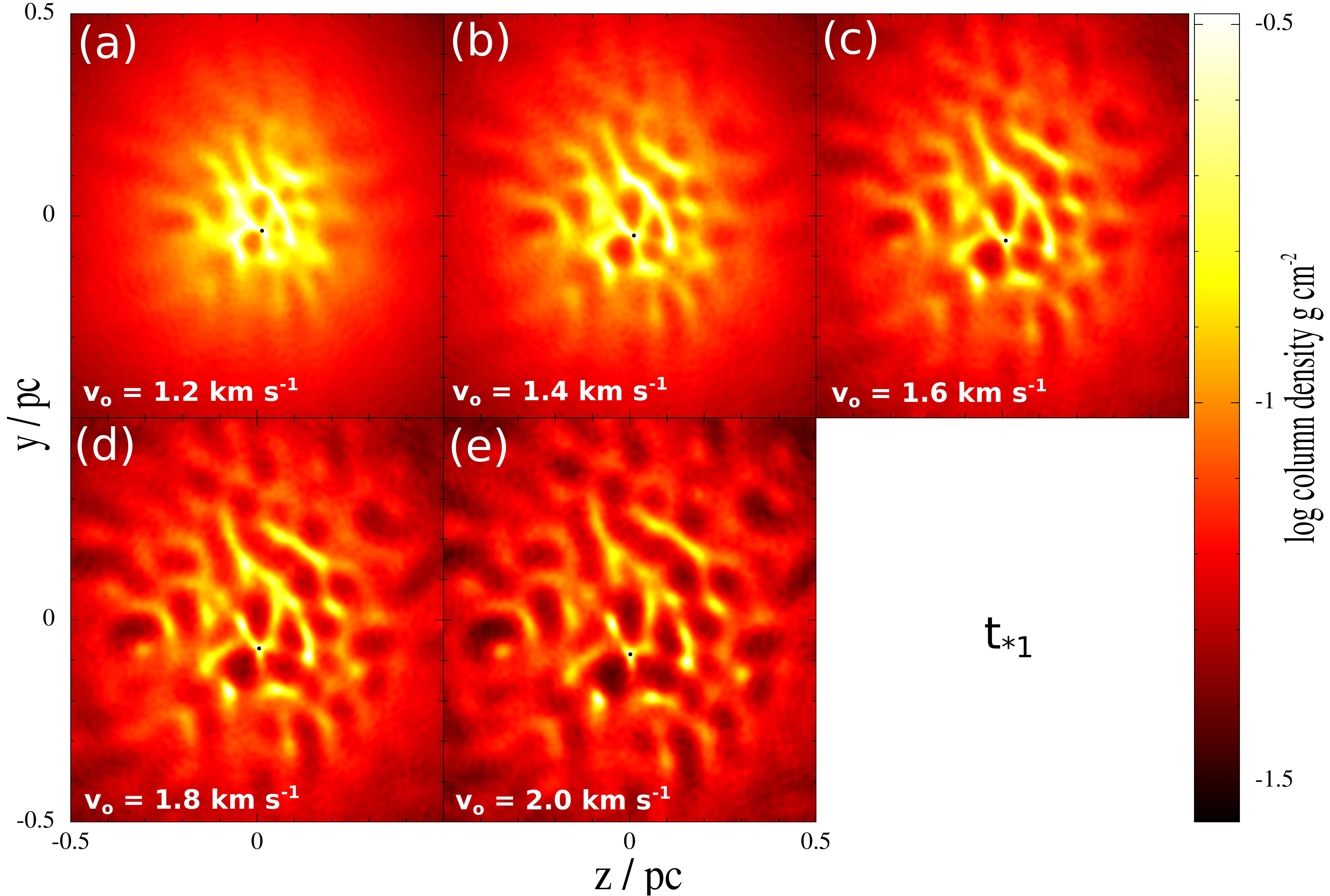}
\caption{False-colour images of fragmenting layers at the moment the first protostar forms, $t_{_{\star 1}}$, from simulations using the same initial clouds (one of the ones that start with weak density and velocity perturbations) but different bulk velocities, $v\subO$. (a) $v\subO\!=\!1.2\,{\rm km}\,{\rm s}^{-1}$, $t_{_{\star 1}}\!=\!1.33\,{\rm Myr}$.  (b) $v\subO\!=\!1.4\,{\rm km}\,{\rm s}^{-1}$, $t_{_{\star 1}}\!=\!1.28\,{\rm Myr}$. (c) $v\subO\!=\!1.6\,{\rm km}\,{\rm s}^{-1}$, $t_{_{\star 1}}\!=\!1.23\,{\rm Myr}$. (d) $v\subO\!=\!1.8\,{\rm km}\,{\rm s}^{-1}$, $t_{_{\star 1}}\!=\!1.19\,{\rm Myr}$. (e) $v\subO\!=\!2.0\,{\rm km}\,{\rm s}^{-1}$, $t_{_{\star 1}}\!=\!1.14\,{\rm Myr}$. Colour represents $\log_{_{10}}\!\left\{\Sigma/{\rm g}\,{\rm cm}^{-2}\right\}$, where $\Sigma$ is column-density projected on the $(y,z)$-plane, i.e. looking along the direction of the collision. The first protostar is represented by a black dot.}
\label{FIG:FaceView@t*1}
\end{figure*}
%%%%%

%%%%%%%
\begin{figure*}
\includegraphics[scale=0.27]{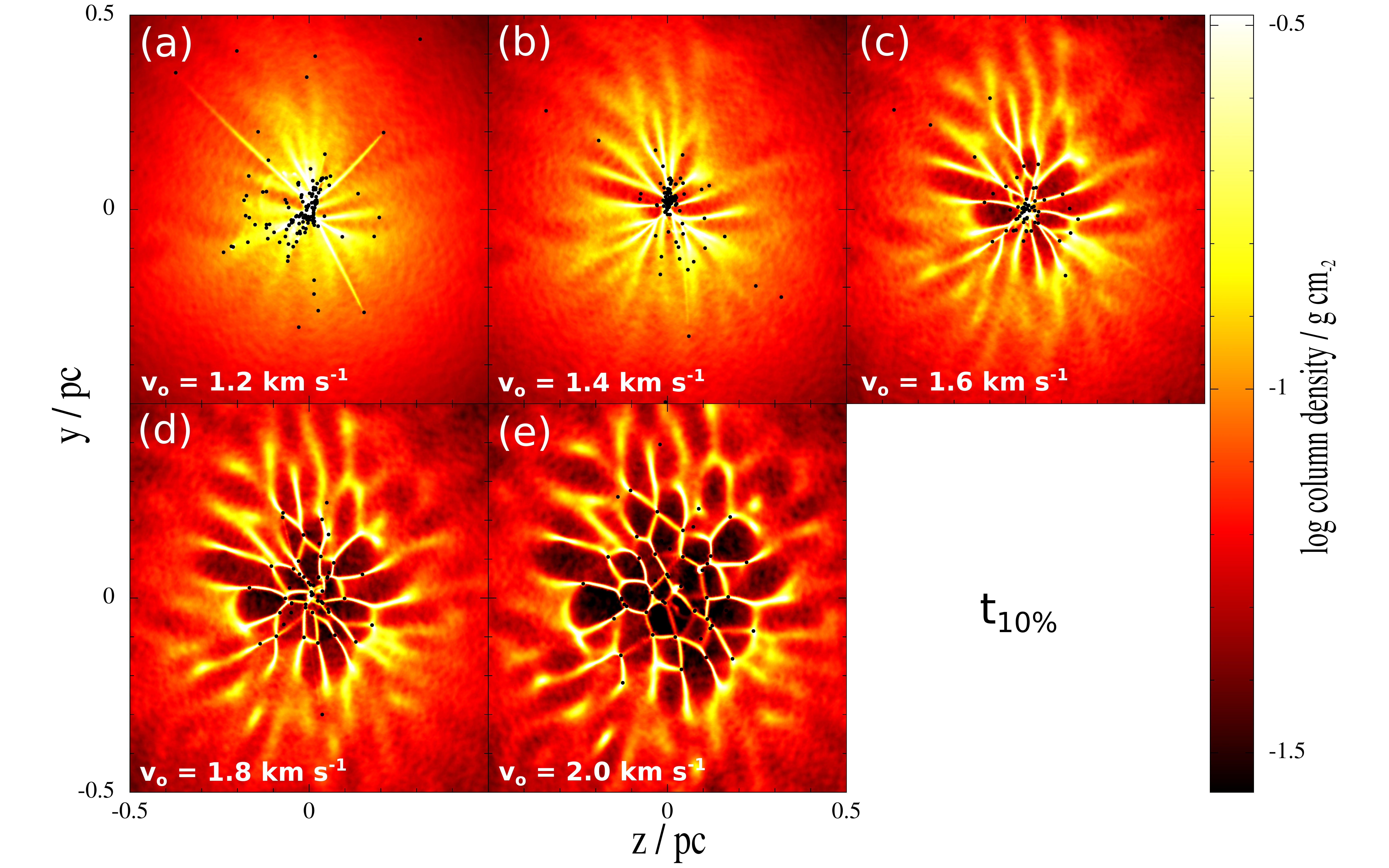}
\caption{False-colour images of fragmenting layers at the moment, $t_{_{10\%}}$, when $10\%$ of the mass has been converted into protostars, from the same simulations presented in Fig. \ref{FIG:FaceView@t*1}. (a) $v\subO\!=\!1.2\,{\rm km}\,{\rm s}^{-1}$, $t_{_{10\%}}\!=\!1.47\,{\rm Myr}$.  (b) $v\subO\!=\!1.4\,{\rm km}\,{\rm s}^{-1}$, $t_{_{10\%}}\!=\!1.42\,{\rm Myr}$. (c) $v\subO\!=\!1.6\,{\rm km}\,{\rm s}^{-1}$, $t_{_{10\%}}\!=\!1.37\,{\rm Myr}$. (d) $v\subO\!=\!1.8\,{\rm km}\,{\rm s}^{-1}$, $t_{_{10\%}}\!=\!1.33\,{\rm Myr}$. (e) $v\subO\!=\!2.0\,{\rm km}\,{\rm s}^{-1}$, $t_{_{10\%}}\!=\!1.28\,{\rm Myr}$. Colour represents $\log_{_{10}}\!\left\{\Sigma/{\rm g}\,{\rm cm}^{-2}\right\}$, where $\Sigma$ is column-density projected on the $(y,z)$-plane, i.e. looking along the direction of the collision. The black dots represent the protostars.}
\label{FIG:FaceView@t10}
\end{figure*}
%%%%%

%%%%%%%%%
\section{Results}\label{SEC:RES}
%%%%%%%%%

%%%%%%%%%%%%
\subsection{Overview}\label{SEC:OVE}
%%%%%%%%%%%%

The cloud-cloud collision creates a dense shock-compressed layer on either side of the $\,x\!=\!0\,$ plane, as illustrated in Fig. \ref{FIG:SideView}. Once this layer becomes sufficiently massive, it fragments gravitationally to produce a network of filaments that feed material into dense prestellar cores, and hence into protostars. At the same time the layer starts to collapse laterally, i.e.  towards the $x$-axis. 

The statistical properties of the protostars that form, and their initial spatial distribution, do not depend on how the initial perturbations are seeded. In other words there is no discernible difference between the simulations of Set 1 (small density perturbations) and those of Set 2 (small density {\it and} velocity perturbations). Therefore we merge the results from the two sets, and discuss them as a single set.

The statistical properties of the protostars that form, and their initial spatial distribution, do depend, critically, on the velocity of the collision, $v\subO$, and we therefore focus on this dependence. We identify two critical times. $\,t_{_{\star 1}}$ is the time at which the first protostar forms (i.e. the first sink particle); Fig. \ref{FIG:FaceView@t*1} presents column-density images at $t_{_{\star 1}}$, for collisions involving the same initial clouds (same initial particle positions and same initial turbulent velocity field) but different bulk velocities, $v\subO$. $\,t_{_{10\%}}$ is the time at which the total mass of protostars first exceeds $100\,{\rm M}_{_\odot}$, in other words $10\%$ of the mass has been converted into protostars; Fig. \ref{FIG:FaceView@t10} presents column-density images at $t_{_{10\%}}$, for the same simulations presented in Fig. {\ref{FIG:FaceView@t*1}. All simulations are terminated at $t_{_{10\%}}$, on the assumption that the inclusion of feedback would decelerate, or even terminate, star formation around this stage -- and also because, by this stage, the progress of a simulation has normally slowed so much that it is not practical to compute much further. 

For lower $v\subO$, the surface-density of the layer builds up more slowly, and therefore there is more time for lateral contraction of the layer (i.e. shrinkage towards the collision axis), before it becomes gravitationally unstable against fragmentation. This leads to a network of filaments that are predominantly radial, and therefore appear like spokes, feeding material into a single, dominant, central hub. Some of the material arriving in the hub has already condensed out as protostars, but some remains as interstellar gas. Consequently, the protostars in the hub compete for this inflowing gas, and one or two grow to very high mass. At the same time, many are ejected by dynamical interactions. By $t_{_{\rm 10\%}}$ there is an extensive diaspora of low-mass protostars, around a single large star cluster.

For higher $v\subO$, the surface-density of the layer builds up more quickly, and so it becomes gravitationally unstable against fragmentation more quickly. As a consequence, the layer has less time to contract laterally, and therefore it fragments into a network of filaments that is more like a spider's web, with many intersections. Prestellar cores condense out at each of these intersections, but, since these cores can only accrete a limited amount of material (i.e. just from the the small sections of filament closest to them), each core spawns only a small sub-cluster, typically comprising one moderately massive protostar and a few lower-mass protostars. Consequently, by $t_{_{10\%}}$ there have been fewer dynamical ejections, and the individual sub-clusters are still relatively isolated from one another. We anticipate that if the evolution were followed further, these sub-clusters would fall together and merge. 

With lower $v\subO$, the maximum stellar mass is larger (than with higher $v\subO$), but the mean mass, and the minimum mass are both lower. With lower $v\subO$, the instantaneous protostellar accretion rates show a greater spread (than with higher $v\subO$), even at high masses, reflecting the stochastic nature of competitive accretion.

\begin{figure}
\includegraphics[scale=0.24]{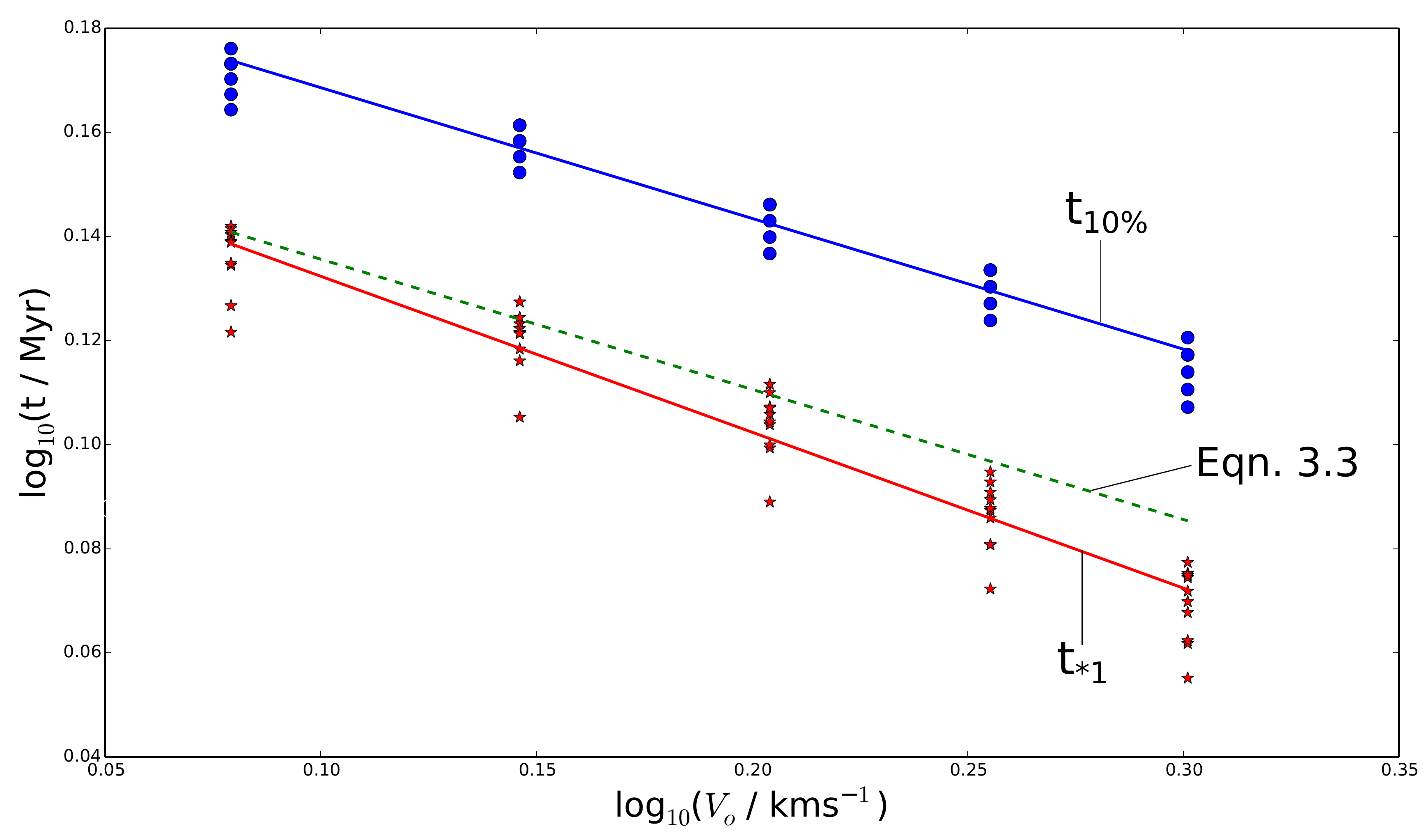}
\caption{The time of formation of the first protostar, $t_{_{\star 1}}$ (red stars, red best-fit line), and the time at which $10\%$ of the total mass has been converted into protostars, $t_{_{10\%}}$ (blue filled circles, blue best-fit line), plotted against the cloud velocity, $v\subO$. The green dashed line is the value of $t_{_{\star 1}}$ predicted by Eqn. (\ref{EQN:t*1}). }
\label{FIG:t*1}
\end{figure}

%%%%%%%%%%%%
\subsection{Fragmentation of the layer}
%%%%%%%%%%%%

Fig. \ref{FIG:FaceView@t*1} shows the filamentary structure that has developed by the time the first protostar forms at $t_{_{\star 1}}$, for simulations with different collision velocities, $v\subO$. In all cases the same initial cloud has been used, in this case a cloud with uniform density and a weak turbulent velocity field. The turbulent velocity field, and the cloud orientation, are the same in all five simulations, and consequently the configuration of the filaments at $t_{_{\star 1}}$ is broadly similar. Indeed, there is a clear predominance of filaments with loci $y\simeq y\subO -1.1z$, reflecting the fact that in this case the initial turbulent velocity field has a significant fraction of its power in a compressional mode with $k_{_z}\simeq +1.1k_{_y}$. This indicates that filaments growth is seeded largely by the initial perturbations, even though these are small (see Section \ref{SEC:PERTS}).

The only difference is that with lower collision velocity the surface-density of the layer builds up more slowly, and so the filamentary structure takes longer to develop; as a result, by the time the first protostar forms, lateral contraction of the layer has progressed further, and the extent of the filamentary network is smaller. 
\subsection{Formation of the first protostar}

A shock-compressed layer created by two semi-infinite anti-parallel flows with density $\rho\subO$ and velocity $v\subO$ fragments at time 
\begin{eqnarray}
t_{_{\rm CRIT}}&\simeq&\left(\frac{a\subS}{2^{3/2}\,G\,\rho\subO\,v\subO}\right)^{1/2}
\end{eqnarray}
(Whitworth 2015, in preparation). Here $a\subS$ is the effective isothermal sound speed in the shocked gas, and `semi-infinite' means that, if the contact discontinuity between the flows is the $x\!=\!0$ plane, the gas has infinite extent in the $y$- and $z$-directions. 

However, the layer created by a cloud-cloud collision has finite extent in the $y$- and $z$-directions, and consequently it contracts towards the $x$-axis on a time-scale
\begin{eqnarray}
t_{_{\rm FF}}&\simeq&\left(\frac{3\,\pi}{32\,G\,\rho\subO}\right)^{1/2}\,,
\end{eqnarray}
thereby accelerating the condensation of the first protostar.

We therefore posit -- somewhat arbitrarily -- that the time at which the first protostar forms might be approximated by the logarithmic mean of $t_{_{\rm CRIT}}$ and $t_{_{\rm FF}}$, i.e.
\begin{eqnarray}\nonumber
t_{_{\star 1}}&\simeq&\left(t_{_{\rm CRIT}}t_{_{\rm FF}}\right)^{1/2}\\\label{EQN:tPF}
&\simeq&\frac{1}{2\,(G\,\rho\subO)^{1/2}}\,\left(\frac{3\,\pi\,a\subS}{2^{5/2}\,v\subO}\right)^{1/4}\,.
\end{eqnarray}\label{EQN:t*1}
Setting $a\subS=0.19\,{\rm km}\,{\rm s}^{-1}$, this reduces to
\begin{eqnarray}
t_{_{\star 1}}&\simeq& 1.45\,{\rm Myr}\,\left(\frac{v\subO}{{\rm km}\,{\rm s}^{-1}}\right)^{-1/4}\,.
\end{eqnarray} 
Fig. \ref{FIG:t*1} shows the values of $t_{_{\star 1}}$ and $t_{_{10\%}}$ from the simulations, plotted against $v\subO$, and the predictions of Eqn. (\ref{EQN:t*1}). 

%%%%%%%%%%%%
\subsection{Reconfiguration of the filaments}
%%%%%%%%%%%%

After $t_{_{\star 1}}$, the network of filaments continues to be amplified, and -- in the cases illustrated in Figs. \ref{FIG:FaceView@t*1} and \ref{FIG:FaceView@t10} -- the predominance of filaments with $k_{_z}\simeq +1.1k_{_y}$ wanes, as other modes in the initial turbulent velocity field assert themselves, to produce a criss-crossing network of filaments. At the same time, the shock-compressed layer contracts laterally. This contraction is non-homologous, because the surface-density of the layer increases towards the $x$-axis, and consequently the inner parts of the layer converge on the centre on a shorter timescale than the outer parts. This has the effect of stretching filaments into predominantly radial orientations. 

At lower collision velocities, $v\subO\la 1.6\,{\rm km}\,{\rm s}^{-1}$, radial stretching is the main process reconfiguring the filaments. This has two consequences. First, the filaments morph into a system of radial arteries feeding matter into the centre, i.e. a hub and spokes system (see the frames on the top row of Fig. \ref{FIG:FaceView@t10}). Second, wherever a filament fragments to produce protostars, these protostars are diverging from one another, due to the tidal stretching of the filament out of which they have just condensed. Therefore they tend to have low masses and they tend not to interact with one another until they reach the central hub. There the protostars interact violently with one another, and most of the low-mass ones are ejected, to produce a diaspora of low-mass protostars, whilst a few settle into the centre and grow by competitive accretion.

At higher collision velocities, $v\subO\ga 1.6\,{\rm km}\,{\rm s}^{-1}$, filaments grow from many different turbulent modes, producing a criss-crossing network, with multiple nodes wherever two or more filaments intersect (see the frames on the bottom row of Fig. \ref{FIG:FaceView@t10}). Each node develops into a prestellar core, but since the amount of matter in the short filamentary segments between nodes is small and is shared between two nodes, these cores normally have quite low mass. Consequently, each individual core can only spawn a small sub-cluster of protostars, and typically this sub-cluster contains just one moderately massive protostar and a few less massive ones.

%%%%%%%%%%%%
\subsection{The rate of star formation}
%%%%%%%%%%%%

The simulations are terminated at $t_{_{10\%}}$, when $10\%$ of the total mass, i.e. $100\,{\rm M}_{_\odot}$, has been converted into protostars. Values of $t_{_{10\%}}$ are plotted against $v\subO$ on Fig. \ref{FIG:t*1}. The duration of star formation,
\begin{eqnarray}
\Delta t_{_{\rm SF}}&=&t_{_{10\%}}-t_{_{\star 1}}\;\,\simeq\;\,0.14\pm0.01\,{\rm Myr}\,,
\end{eqnarray}
is essentially independent of the collision velocity, $v\subO$. Thus, the overall star formation rate in the layer is also essentially independent of $v\subO$,
\begin{eqnarray}
\dot{M}_{_\star}&=&\frac{100\,{\rm M}_{_\odot}}{\Delta t_{_{\rm SF}}}\;\,\simeq\;\,700\,{\rm M}_{_\odot}\,{\rm Myr}^{-1}\,.
\end{eqnarray}
At lower collision velocity, star formation starts later, but there is less mass in the shock-compressed layer, when it fragments. This is compensated by the fact that the layer has had more time to contract laterally, so the surface-density of the layer is comparable, and hence the star formation rate is similar.

%%%%%%%
\begin{figure}
\includegraphics[scale=0.24]{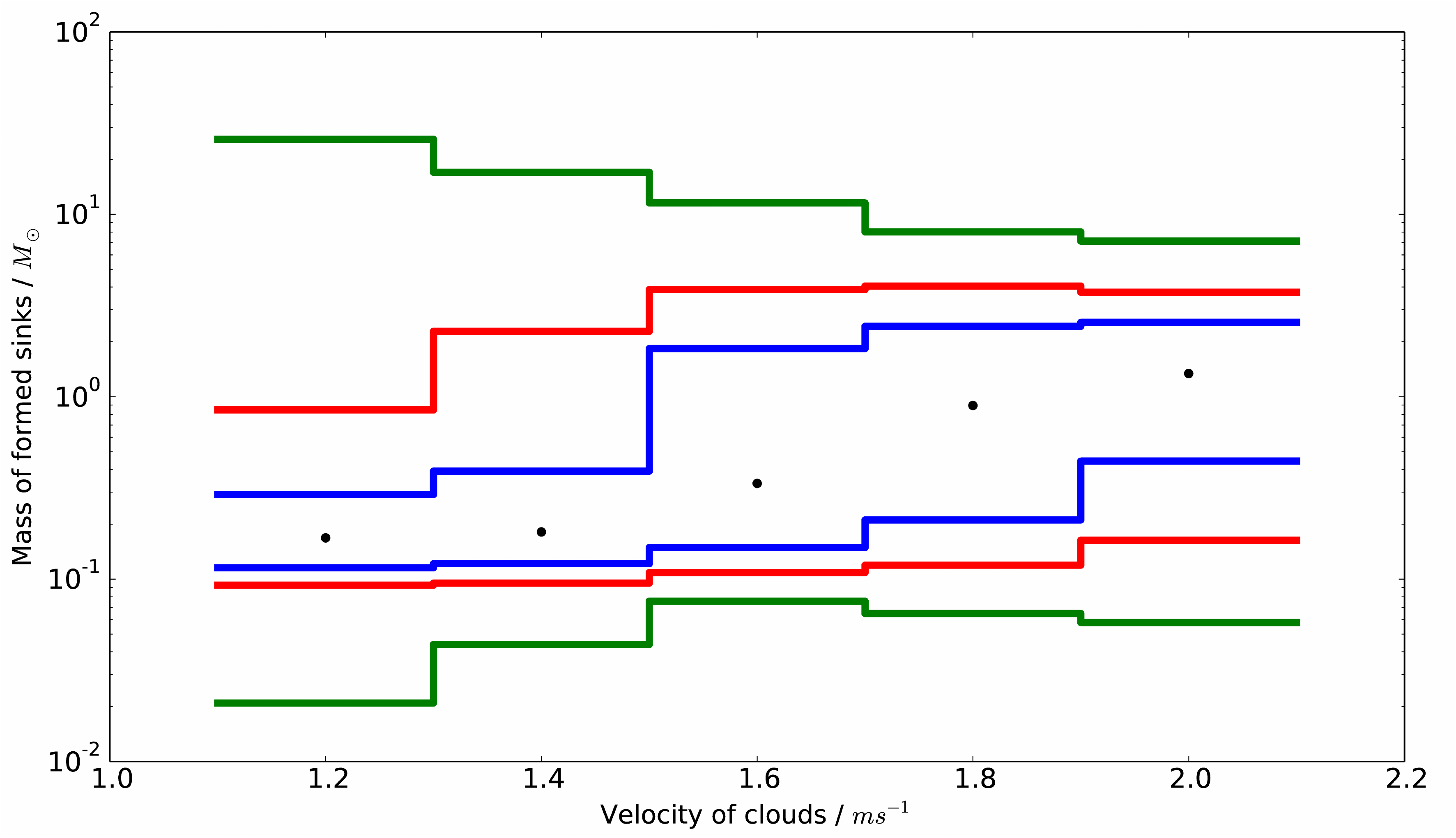}
\caption{Protostellar mass ranges. For each collision velocity, $v\subO$, we combine the data from all ten realisations. { The filled circles give the median mass. Scanning downwards (upwards) from the median mass, the first {\it blue} line gives the 25$^{\rm th}$ (75$^{\rm th}$) centile, i.e. the interquartile range; the next {\it red} line gives the 10$^{\rm th}$ (90$^{\rm th}$) centile; and the last {\it green} line gives the 0$^{\rm th}$ (100$^{\rm th}$) centile (i.e. the full mass range).}}
\label{FIG:PercentileRanges}
\end{figure}
%%%%%

%%%%%%%
\begin{figure}
\includegraphics[scale=0.24]{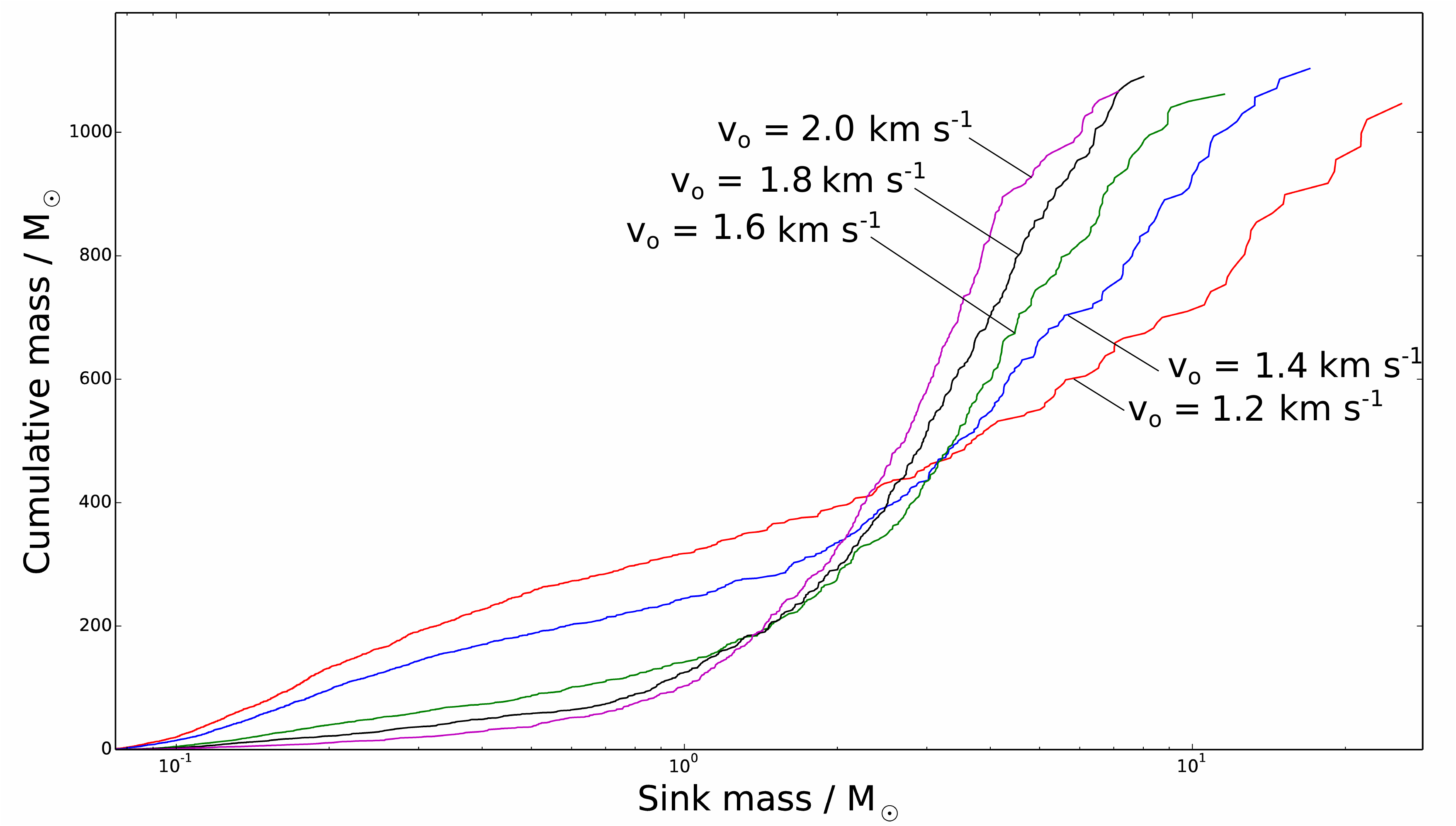}
\caption{The cumulative mass of protostars at $t_{_{10\%}}$, as a function of protostellar mass: red curve, $v\subO\!=\!1.2\,{\rm km}\,{\rm s}^{-1};\;$ blue curve, $v\subO\!=\!1.4\,{\rm km}\,{\rm s}^{-1};\;$ green curve, $v\subO\!=\!1.6\,{\rm km}\,{\rm s}^{-1};\;$ black curve, $v\subO\!=\!1.8\,{\rm km}\,{\rm s}^{-1};\;$ purple curve, $v\subO\!=\!2.0\,{\rm km}\,{\rm s}^{-1}$.}
\label{FIG:Cumulative}
\end{figure}
%%%%%

%%%%%%%
\begin{table}
\caption{Parameters of the mass functions of protostars formed with different collision velocities. Column 1 gives the collision velocity, $v\subO$, column 2 gives the mean number of protostars, $\bar{\cal N}_{_\star}$, formed from one collision (and totalling $100\,{\rm M}_{_\odot}$), and columns 3 through 6 give, respectively, the mean, $\mu\subL$, standard deviation, $\sigma\subL$, skew, $\gamma\subL$, and kurtosis, $\kappa\subL$, of the protostellar log-masses, $\log_{_{10}}\!\!\left(\!M_{_\star}/{\rm M}_{_\odot}\!\right)$}
\begin{center}
\begin{tabular}{crcccc}\hline
$v\subO$ & $\bar{\cal N}_{_\star}$ & $\mu\subL$ & $\sigma\subL$ & $\gamma\subL$ & $\kappa\subL$ \\\hline
$1.2\,{\rm km}\,{\rm s}^{-1}$ & 198 & -0.65 & +0.45 & +1.92 & +4.08 \\
$1.4\,{\rm km}\,{\rm s}^{-1}$ & 163 & -0.56 & +0.53 & +1.41 & +1.12 \\
$1.6\,{\rm km}\,{\rm s}^{-1}$ & 108 & -0.29 & +0.60 & +0.43 & -1.21 \\
$1.8\,{\rm km}\,{\rm s}^{-1}$ & 88 & -0.13 & +0.58 & -0.11 & -1.33 \\
$2.0\,{\rm km}\,{\rm s}^{-1}$ & 86 & -0.01 & +0.50 & -0.58 & -0.72 \\\hline
\end{tabular}
\end{center}
\label{TAB:MF}
\end{table}
%%%%%

%%%%%%%%%%%%
\subsection{The mass function of the protostars}
%%%%%%%%%%%%

As the collision velocity, $v\subO$, is increased, the median protostellar mass increases, but the maximum protostellar mass and the spread of protostellar masses both decrease. Lower $v\subO$ leads to the formation of a single central hub, into which the majority of protostars fall; there they partake in competitive accretion, and consequently a small number of protostars grow to very large masses, and a large number of protostars are ejected dynamically, producing an extensive diaspora of predominantly low-mass protostars. At higher $v\subO$, there are many different cores spawning protostars independently of one another. In each such core a moderately massive protostar forms, along with a few somewhat lower-mass protostars; in these small-$N$ sub-clusters, dynamical ejections do not occur so quickly, and when the simulations are terminated at $t_{_{10\%}}$ most protostars are still near the location where they were born.

Fig. \ref{FIG:PercentileRanges} illustrates these trends by plotting the mass ranges, as a function of collision velocity. The green lines represent the complete range of masses at each velocity. The upper green line here is significant, and reflects the fact that the most massive protostars are formed in the lower-$v\subO$ collisions. The lower green line simply reflects a single very low-mass protostar formed in one of the ten simulations at that particular collision velocity, and is therefore less significant. The red lines represent the range between the 10$^{\rm th}$ and 90$^{\rm th}$ centiles; in other words, {\it by number}, $10\%$ of stars fall below this line, and $10\%$ fall above it. Similarly, the blue lines represent the range between the 25$^{\rm th}$ and 75$^{\rm th}$ centiles, also known as the interquartile range. The filled circles represent the median masses for each collision velocity.

Fig. \ref{FIG:Cumulative} shows the cumulative mass of protostars at $t_{_{10\%}}$, as a function of protostellar mass, at the different collision velocities. This illustrates how the number of low-mass protostars increases dramatically with decreasing collision velocity, $v\subO$, once $v\subO$ falls below $\,\sim 1.6\,{\rm km}\,{\rm s}^{-1}$. It also illustrates how the two most massive protostars become more massive with decreasing $v\subO$ as $v\subO$ falls below $\,\sim 1.6\,{\rm km}\,{\rm s}^{-1}$. For higher collision velocities, $v\subO \ga 1.6\,{\rm km}\,{\rm s}^{-1}$, the distribution of masses is much tighter, i.e. it has a smaller standard deviation (see Table \ref{TAB:MF}).

In no case is the distribution of masses log-normal. Table \ref{TAB:MF} gives the mean, $\mu\subL$, standard deviation, $\sigma\subL$, skew, $\gamma\subL$, and kurtosis, $\kappa\subL$, of $\log_{_{10}}\!\!\left(\!M_{_\star}/{\rm M}_{_\odot}\!\right)$, for different $v\subO$. As $v\subO$ increases, $\mu\subL$ increases monotonically, i.e. the mean stellar mass increases; $\sigma\subL$ increases and then decreases, but does not vary by much $(0.45\la\sigma\subL\la 0.60)$; $\gamma\subL$ decreases monotonically, indicating that the mass function is very assymetric at low $v\subO$ (due to the large number of very low-mass protostars), but becomes increasingly symmetric at larger $v\subO$; the variation of the kurtosis indicates that the mass function at low $v\subO$ is much more sharply peaked than at high $v\subO$.

Although on large scales the {\it observed} Initial Mass Function (IMF) appears to vary very little, it is still possible that there exists some scale below which there are significant variations, over and above those that arise purely from sampling statistics; the observed universal IMF then constitutes an average over a representative ensemble of such regions. The results above are a prediction for what might be uncovered if the stellar population resulting from a single cloud-cloud collision could be accurately collated, at an early stage, before it is dispersed and/or contaminated with other stars.

%%%%%%%
\begin{figure*}
\includegraphics[scale=0.47]{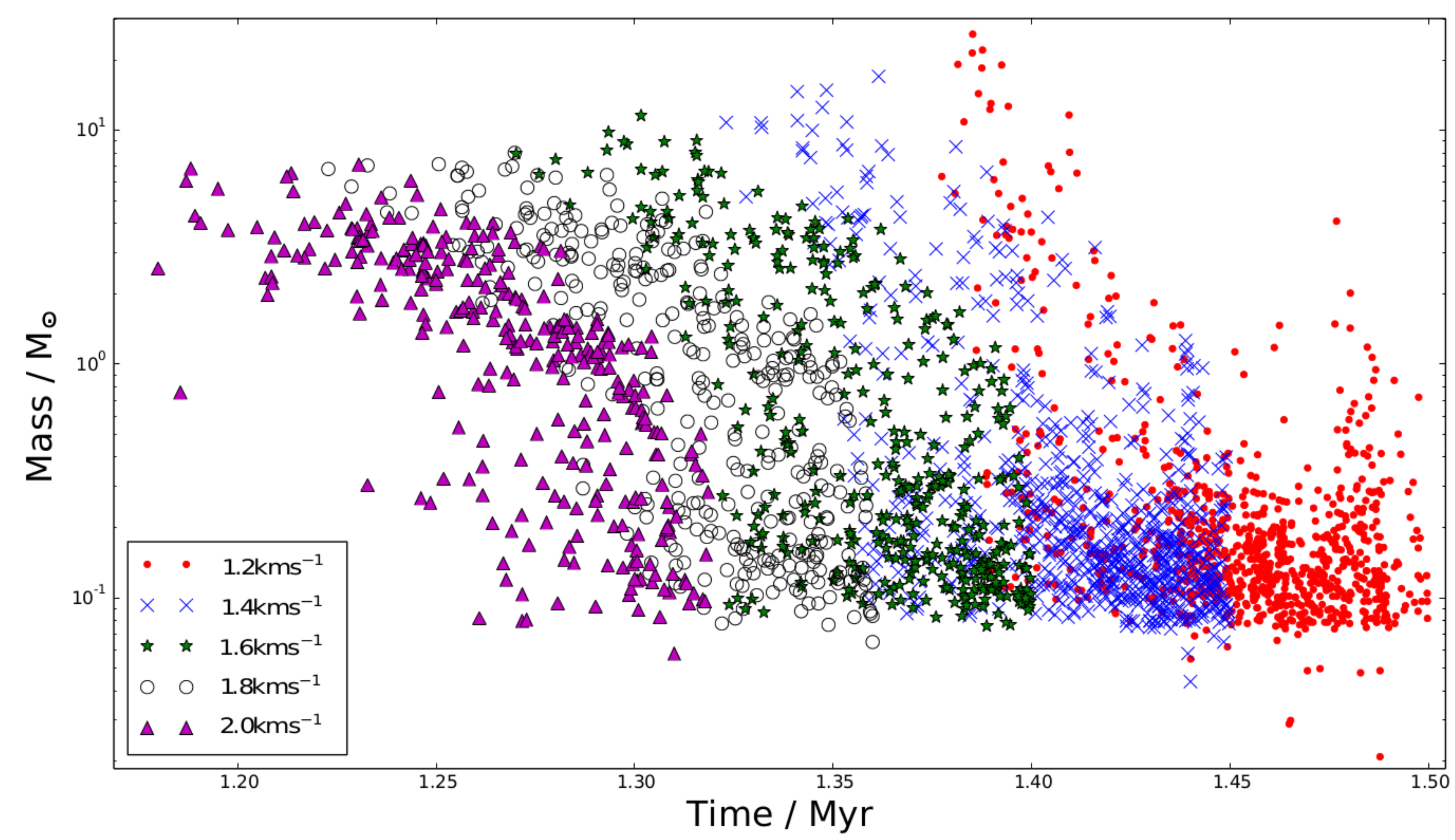}
\caption{The final mass of each protostar, $M_{_\star}$, at $t_{_{\rm 10\%}}$ (the end of the simulation) plotted against its time of formation, $t_{_{\rm FORM}}$, i.e. the time at which the corresponding sink particle is created. Red dots, $v\subO\!=\!1.2\,{\rm km}\,{\rm s}^{-1};\;$ blue crosses, $v\subO\!=\!1.4\,{\rm km}\,{\rm s}^{-1};\;$ green stars, $v\subO\!=\!1.6\,{\rm km}\,{\rm s}^{-1};\;$ black open circles, $v\subO\!=\!1.8\,{\rm km}\,{\rm s}^{-1};\;$ purple triangles, $v\subO\!=\!2.0\,{\rm km}\,{\rm s}^{-1}$.}
\label{FIG:MFvsTF}
\end{figure*}
%%%%%

%%%%%%%
\begin{figure}
\includegraphics[width=\columnwidth]{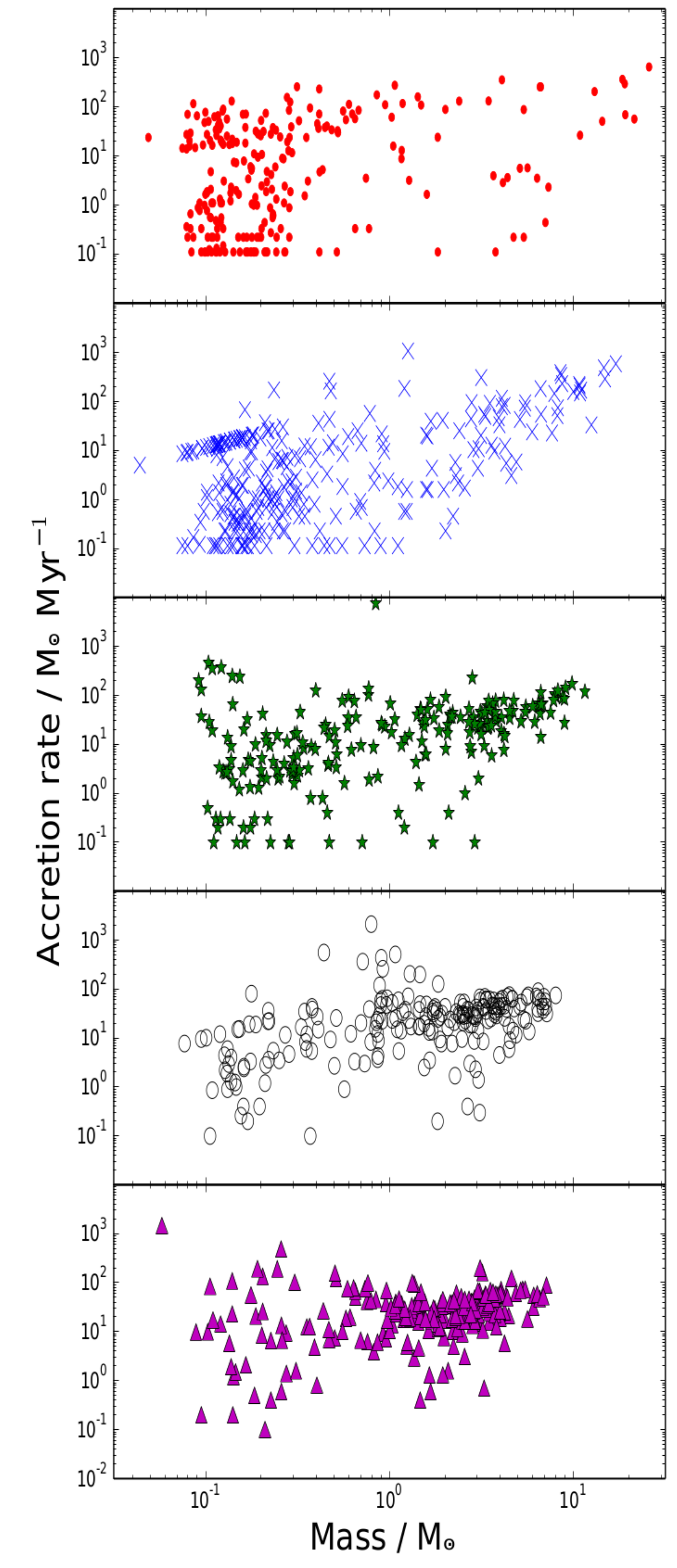}
\caption{Log-log plot of the rate of accretion, $\dot{M}_{_\star}$, against the protostellar mass, $M_{_\star}$, both evaluated at the end of the simulation, $t_{_{10\%}}$. Red dots, $v\subO\!=\!1.2\,{\rm km}\,{\rm s}^{-1};\;$ blue crosses, $v\subO\!=\!1.4\,{\rm km}\,{\rm s}^{-1};\;$ green stars, $v\subO\!=\!1.6\,{\rm km}\,{\rm s}^{-1};\;$ black open circles, $v\subO\!=\!1.8\,{\rm km}\,{\rm s}^{-1};\;$ purple triangles, $v\subO\!=\!2.0\,{\rm km}\,{\rm s}^{-1}$. Note that the lowest accretion rates are quantised at values $0.1\,{\rm M}_{_\odot}\,{\rm Myr}^{-1}$, $0.2\,{\rm M}_{_\odot}\,{\rm Myr}^{-1}$, $0.3\,{\rm M}_{_\odot}\,{\rm Myr}^{-1}$, etc., corresponding to sinks accreting one, two, three, etc. SPH particles (hence $10^{-3}\,{\rm M}_{_\odot}$, $2\times 10^{-3}\,{\rm M}_{_\odot}$ $3\times 10^{-3}\,{\rm M}_{_\odot}$, etc.), during the $0.01\,{\rm Myr}$ interval between the last two outputs. Similarly some accretion rates fall on the line $\dot{M}_{_\star}=M_{_\star}/0.01\,{\rm Myr}$, corresponding to sinks formed during the $0.01\,{\rm Myr}$ interval between the last two outputs.}
\label{FIG:MdotF}
\end{figure}
%%%%%

%%%%%%%%%%%%
\subsection{Accretion rates}
%%%%%%%%%%%%
Fig. \ref{FIG:MFvsTF} shows the final protostellar masses, $M_{_\star}$, at $t_{_{10\%}}$, plotted against their time of formation, $t_{_{\rm FORM}}$, i.e. the time at which the corresponding sink particle was created. As the collision velocity is increased, the formation time time decreases, the maximum mass decreases, and the number of low-mass protostars decreases.

Fig. \ref{FIG:MdotF} shows the protostellar accretion rates, $\dot{M}_{_\star}$, plotted against protostellar mass, $M_{_\star}$, both evaluated at the end of the simulation, $t_{_{10\%}}$. For lower collision velocities, $v\subO <1.6\,{\rm km}\,{\rm s}^{-1}$, final masses are concentrated towards low values, $0.08\;{\rm to}\;3\,{\rm M}_{_\odot}$, and there is a wide range of final accretion rates at all masses, $10^{-1}\;{\rm to}\;10^2\,{\rm M}_{_\odot}\,{\rm Myr}^{-1}$, apart from the most massive protostars, which are all still growing very rapidly. For higher collision velocities, $v\subO > 1.6\,{\rm km}\,{\rm s}^{-1}$, final masses and final accretion rates are concentrated at quite high values, $1\;{\rm to}\;10\,{\rm M}_{_\odot}$ and $10\;{\rm to}\;100\,{\rm M}_{_\odot}\,{\rm Myr}^{-1}$, respectively.

%%%%%%%
\begin{figure}
\includegraphics[width=\columnwidth]{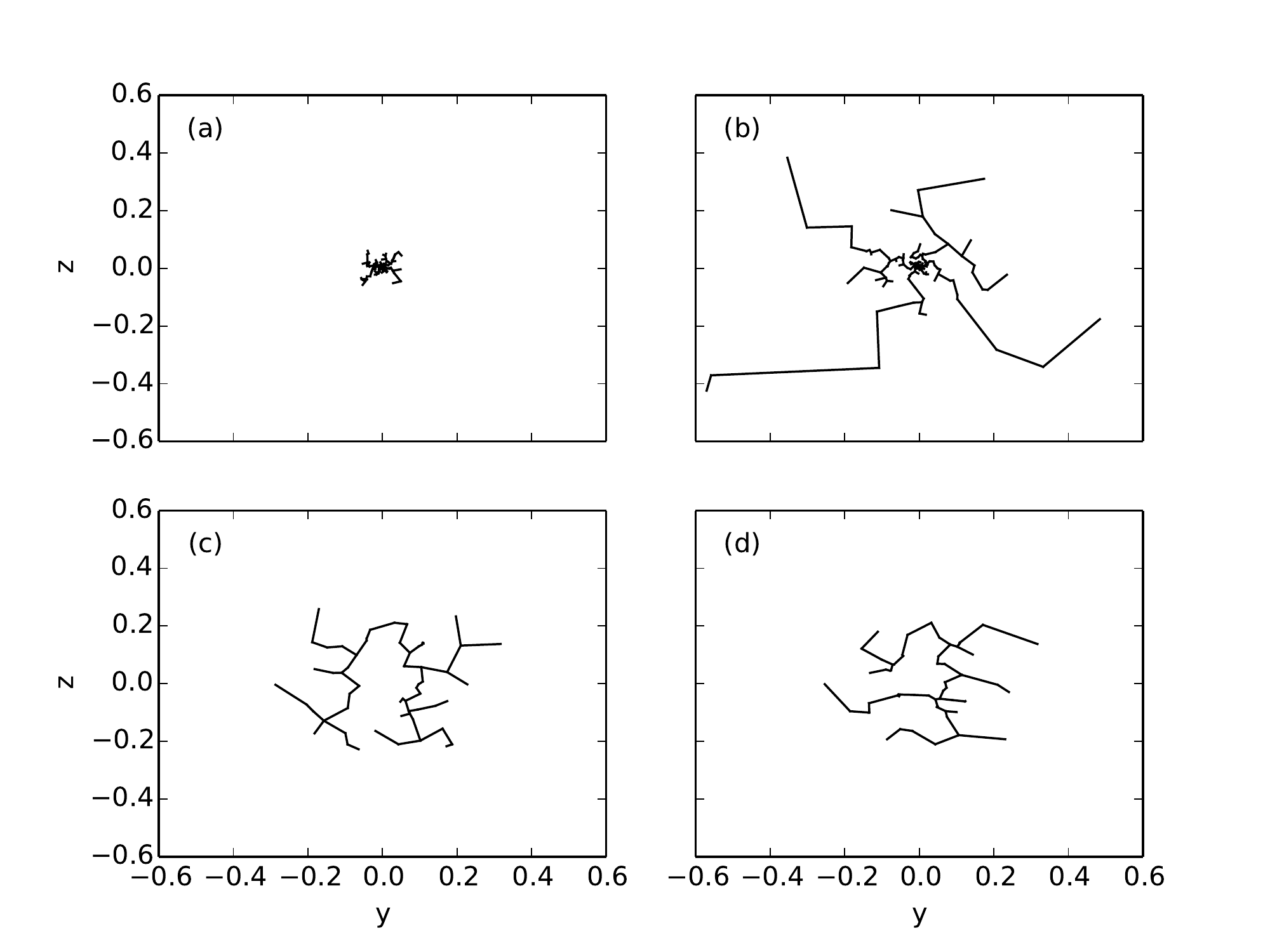}
\caption{Minimal Spanning Trees connecting the positions of protostars projected on the $x\!=\!0$ plane (the collision axis). The top row represents a simulation with $v\subO=1.2\,{\rm km}\,{\rm s}^{-1}$; (a) shows the initial positions (i.e. where the protostars formed), and (b) shows the final positions (i.e. at $t_{_{10\%}}$). The bottom row represents a simulation with $v\subO=2.0\,{\rm km}\,{\rm s}^{-1}$; (c) shows the initial positions, and (d) the final positions.}
\label{FIG:MST}
\end{figure}
%%%%%

%%%%%%%%%%%%
\subsection{Spatial distribution of protostars}
%%%%%%%%%%%%

We have analysed the spatial distributions of the protostars formed in the simulations, and how they evolve, by projecting the initial and final positions of protostars onto the $x\!=\!0$ plane and constructing Minimal Spanning Trees (MSTs). The top row of Fig. \ref{FIG:MST} shows an example of the MSTs for the initial (left) and final (right) positions obtained in a low-velocity collision, with $v\subO=1.2\,{\rm km}\,{\rm s}^{-1}$. The bottom row of Fig. \ref{FIG:MST} shows an example of the MSTs for the initial (left) and final (right) positions obtained in a high-velocity collision, with $v\subO=2.0\,{\rm km}\,{\rm s}^{-1}$. We see that in the aftermath of a low-velocity collision the initial positions of the protostars (Fig. \ref{FIG:MST}a) are strongly concentrated in a central cluster, but the final positions (Fig. \ref{FIG:MST}b) are spread out over a large area, due to violent dynamical ejections of low-mass protostars. In contrast, in the aftermath of a high-velocity collision the initial positions (Fig. \ref{FIG:MST}c) are quite spread out, and the final positions (Fig. \ref{FIG:MST}d) are, statistically, not much changed, reflecting the fact that thus far there has been little dynamical evolution within the separate sub-clusters, and the protostars are still quite close to the places where they formed.

This is quantified in Fig. \ref{FIG:EDGEPDF}, where we plot the mean edge-lengths on the initial and final MSTs against the collision velocity, $v\subO$. The mean edge-length on the {\it initial} MST is a strong function of the collision velocity, increasing by almost an order of magnitude (lower blue curve), as we go from $v\subO=1.2\,{\rm km}\,{\rm s}^{-1}$ (very concentrated single cluster) to $v\subO=2.0\,{\rm km}\,{\rm s}^{-1}$ (many more-or-less independent sub-clusters). The mean edge-length on the {\it final} MST is almost independent of the collision velocity (upper green curve). For lower-velocity collisions, this is largely due to the dispersal of the low-mass protostars ejected from the central cluster. As the collision velocity is increased, this effect becomes increasingly unimportant, and the mean edge-lengths on the initial and final MSTs obtained with $v\subO\!=\!2.0\,{\rm km}\,{\rm s}^{-1}$ are indistinguishable.

%%%%%%%
\begin{figure}
\includegraphics[width=\columnwidth]{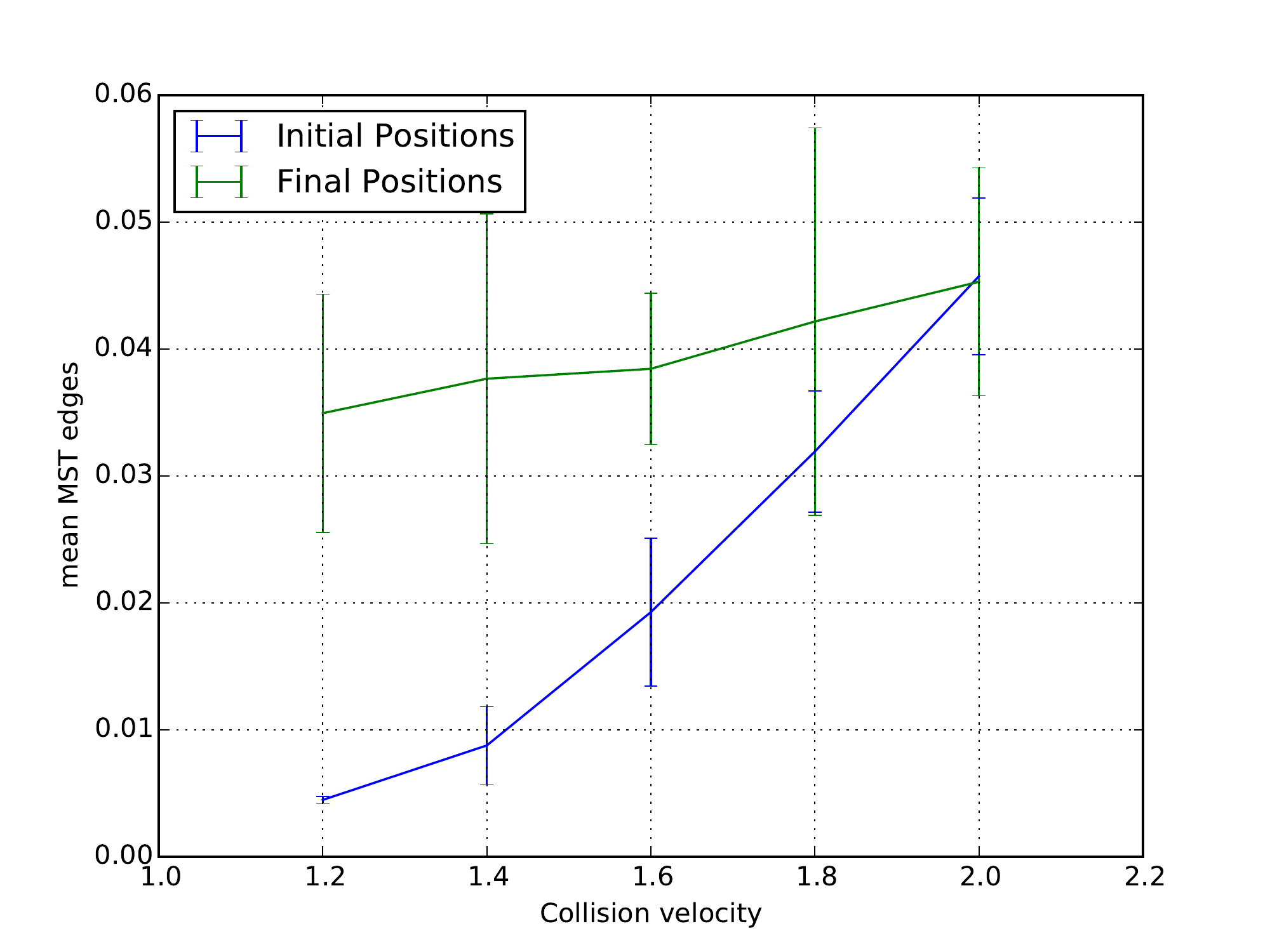}
\caption{The mean edge-lengths from the initial and final MSTs, as a function of collision velocity, $v\subO$. The lower blue curve represents the initial MSTs, and the upper green curve represents the final MSTs.}
\label{FIG:EDGEPDF}
\end{figure}
%%%%%

%%%%%%%
\begin{figure}
\includegraphics[width=\columnwidth]{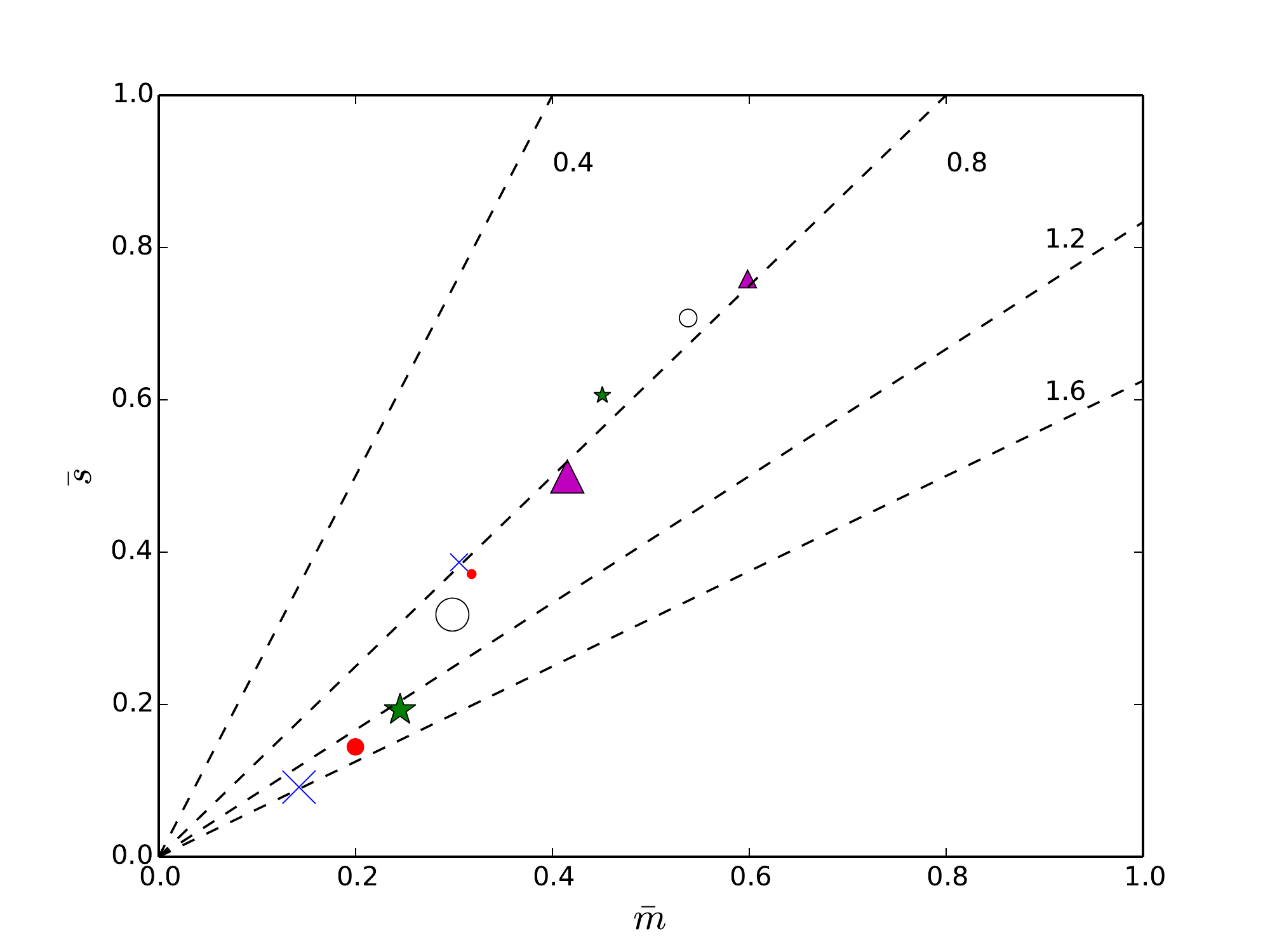}
\caption{The locus of simulations on the $(\bar{m},\bar{s})$-plane, where $\bar{m}$ and $\bar{s}$ are defined in Eqns. (\ref{EQN:mBAR}) and (\ref{EQN:sBAR}). Red dots mark the mean for $v\subO\!=\!1.2\,{\rm km}\,{\rm s}^{-1};\;$ blue crosses, for $v\subO\!=\!1.4\,{\rm km}\,{\rm s}^{-1};\;$ green stars, for $v\subO\!=\!1.6\,{\rm km}\,{\rm s}^{-1};\;$ black open circles, for $v\subO\!=\!1.8\,{\rm km}\,{\rm s}^{-1};\;$ and purple triangles, for $v\subO\!=\!2.0\,{\rm km}\,{\rm s}^{-1}$. For each $v\subO$ the symbol is smaller for the initial positions and large for the final positions. The dashed lines are lines of constant ${\cal Q}=\bar{m}/\bar{s}$ (Cartwright \& Whitworth 2004).}
\label{FIG:Q}
\end{figure}
%%%%%

In order to estimate the ${\cal Q}$-parameter, and hence quantify further the evolution of the clusters, we must normalise the edge-lengths on the MST according to
\begin{eqnarray}\label{EQN:mBAR}
\bar{m}&=&\frac{(\pi{\cal N}_{_\star})^{1/2}}{({\cal N}_{_\star}-1)}\,\sum\limits_{\rm MST}^{}\,\left\{m_{_{\rm MST}}\right\}\,,
\end{eqnarray}
where ${\cal N}_{_\star}$ is the total number of protostars and $m_{_{\rm MST}}$ is one of the edges on the MST (see Cartwright \& Whitworth 2004). In addition, we must evaluate the mean of the edges connecting all pairs of protostars,
\begin{eqnarray}\label{EQN:sBAR}
\bar{s}&=&\frac{2}{{\cal N}_{_\star}({\cal N}_{_\star}\!\!-1)}\;\,\sum\limits_{i=1}^{i={\cal N}_{_\star}-1}\;\sum\limits_{j=i+1}^{j={\cal N}_{_\star}}\,\left\{s_{ij}\right\}\,,
\end{eqnarray}
where $s_{ij}$ is the edge between protostars $i$ and $j$. Then the ${\cal Q}$-parameter is given by
\begin{eqnarray}
{\cal Q}&=&\frac{\bar{m}}{\bar{s}}\,.
\end{eqnarray}

Fig. \ref{FIG:Q} shows the mean location on the $(\bar{m},\bar{s})$-plane for the clusters produced with different $v\subO$. For each value of $v\subO$, the symbol representing the initial positions of the protostars is smaller than the symbol representing the final positions of the protostars. The dashed lines are lines of constant ${\cal Q}$. In all cases, the cluster evolves towards smaller $\bar{m}$ and smaller $\bar{s}$, but larger ${\cal Q}$. This reflects two factors. First, locally protostars are becoming more tightly bound to their nearest neighbours, which makes the mean edge-lengths (both $\bar{m}$ and $\bar{s}$) smaller. Second, the clusters formed following lower-velocity collisions are strongly centrally condensed, and become even more so, as the higher-mass protostars settle into the centre, and the lower-mass protostars are ejected at high speed; conversely, the clusters formed following higher-velocity collisions are made up of many more-or-less independent sub-clusters, formed at the intersections of a spider's web of filaments, and at the end these sub-clusters are only just beginning to dissolve. 

%%%%%%%%%%%%
\subsection{More extreme collision velocities}
%%%%%%%%%%%%

{ If the clouds considered here $(500{\rm M}_{_\odot},2{\rm pc})$ have lower velocities than we have discussed, $v\subO\la 1\,{\rm km}\,{\rm s}^{-1}$, they collapse individually (and form separate star clusters) faster than they collide, so the effect of the collision is simply to merge two existing star clusters. Conversely, if the clouds have higher velocities than we have discussed, $v\subO\ga 2.4\,{\rm km}\,{\rm s}^{-1}$, the trends described above continue: a more extended spider's web of intersecting filaments forms; the spacing of the filaments decreases, and hence the masses of the subclusters forming at their intersections decreases; star formation occurs earlier; the median mass tends to a constant value; and the range of masses continues to decrease. We have tested this trend up to $v\subO =4\,{\rm km}\,{\rm s}^{-1}$, $\Delta v\subO =8\,{\rm km}\,{\rm s}^{-1}$.}

%%%%%%%%%
\section{Conclusions}\label{SEC:CONC}
%%%%%%%%%

We have explored the phenomenology of star formation triggered by cloud/cloud collisions, specifically what happens when two quiescent and  approximately uniform-density clouds with mass $500\,{\rm M}_{_\odot}$ and radius $2\,{\rm pc}$ collide head-on at relative speeds between $2.4$ and $4.0\,{\rm km}\,{\rm s}^{-1}$.
\begin{enumerate}
\item{In all cases the colliding clouds produce a shock-compressed layer, which fragments into filaments.}
\item{At the lower relative velocities, it takes longer before fragmentation becomes non-linear and delivers protostars.}
\item{At the lower relative velocities the lateral collapse of the layer drags and stretches the filaments towards the collision axis, so that they become predominantly radial. Protostars that condense out of the filaments fall along the filaments into the centre, along with residual gas, and a single massive cluster forms there. Competitive accretion leads to the growth of a few very massive protostars, and the ejection of a large number of low-mass protostars. The distribution of filaments around the single massive cluster can be described as a hub and spokes system.}
\item{At the higher relative velocities, there is less time for lateral collapse of the layer, and so the filaments form a network like a spider's web, with many more-or-less independent star-forming cores collecting at the intersections of the filaments. Since these cores grow from relatively short filamentary segments, they have limited mass, and typically spawn only a small number of protostars, usually one fairly massive one and a few smaller ones. Because there is less mass, there are fewer protostars in each core, so dynamical ejections occur later and are less violent.}
\item{As the collision velocity is increased, the mean protostellar mass increases, but the range of protostellar masses goes down, so the maximum mass actually decreases and the minimum mass increases; there are many fewer low-mass protostars.}
\item{The spatial distribution of protostars when they first form is marginally fractal, due to the underlying filamentary distribution of the gas. At lower collision velocities, the protostars quickly relax to a single very centrally condensed configuration. At higher collision velocities, the protostars are still in small independent sub-clusters, close to where they formed, at the end of the simulation.}
\end{enumerate}

%%%%%%%%%%
\section*{Acknowledgements}
%%%%%%%%%%

SKB gratefully acknowledges the support of a PhD studentship, and APW the support of a consolidated grant (ST/K00926/1), from the UK Science and Technology Funding Council. DAH acknowledges the the support of the DFG cluster of excellence `Origin and Structure of the Universe'. We would like to thank Dr. J. Palous for his very useful comments which helped to improve this paper. This work was performed using the computational facilities of the Advanced Reaserch Computing at Cardiff (ARCCA) Division, Cardiff Unviersity. All false-colour images have been rendered with SPLASH \citep{Price2007f}.

%%%%%%%%%%%%%%%%%%%%%%%%
\bibliographystyle{mn2e}
\bibliography{ref}
%%%%%%%%%%%%%%%%%%%%%%%

\label{lastpage}
\end{document}